\documentstyle[12pt,epsfig,wrapfig]{article}
\include{times,mathptm}

\renewcommand{\baselinestretch}{1.02}
\newcommand{\degr}{^\circ}
\renewcommand{\section}[1]{\vspace{6pt} \noindent\mbox{#1} \newline
\noindent}
\renewcommand{\subsection}[1]{\vspace{6pt} \noindent\mbox{\underline{#1}}
\newline \noindent}
\renewcommand{\subsubsection}[1]{\vspace{6pt}
\noindent\mbox{\underline{#1}}
\noindent}
\renewcommand\figurename{\it Fig.}
\newfont{\sansb}{cmssbx10}
\newfont{\sans}{cmss10}
\newcommand\tensor[1]{\mbox{\sansb #1}}
\newcommand\vect[1]{\mbox{\bf #1}}
\newcommand\dd{\partial}
\newcommand\sun[0]{\odot}
\newcommand\llyn[0]{\underline{\hspace{1.4cm}}}

\begin{document}

{\small HE 4.1.27 \vspace{-24pt}\\}     
{\center \LARGE {\bf SADCO: HYDROACOUSTIC DETECTION OF
SUPER-HIGH ENERGY COSMIC NEUTRINOS}
\vspace{6pt}\\}
L. G. Dedenko$^1$, A. V. Furduev$^2$, Ya. S. Karlik$^3$, J. G.
Learned$^4$,\\ A. A. Mironovich$^1$, V. D. Svet$^2$ , and I. M.
Zheleznykh$^1$
\vspace{6pt}\\
{\it $^1$ Institute for Nuclear Research of Russian Academy of Sciences,
Moscow, Russia\\
$^2$ N. N. Andreev Acoustic Institute, Moscow, Russia\\
$^3$ Central Research Institute ``Morfizpribor'', St.Petersburg, Russia\\
$^4$ Department of Physics and Astronomy, University of Hawaii, USA
\vspace{-12pt}\\}
{\center ABSTRACT\\}
An attractive technique to explore for super-high-energy cosmic
neutrino fluxes, via deep underwater acoustic detection, is
discussed. Acoustic signals emitted by the neutrino induced cascades
at large distances (10-50 km) from cascades are considered. It is
argued that an existing hydroacoustic array of 2400 hydrophones,
which is available in the Great Ocean near Kamchatka Peninsula,
could be used as a base for an exploratory acoustic neutrino
telescope SADCO (Sea Acoustic Detector of Cosmic Objects). The
detection volume for registration of cascades with energies
in the range of $10^{20-21} eV$ is estimated to be hundreds of cubic
kilometers. Some models of extremely high energy elementary particle
production in the Universe (for example the topological defect
model) may be examined by such a detector.  Tests of this technique
are hoped for within a year.

\setlength{\parindent}{1cm}
\section{INTRODUCTION}
During the last several decades a new branch of physics and
astrophysics, namely high energy neutrino astrophysics (or neutrino
astronomy), has been pursued aggressively by a relatively small but
dedicated community of physicists. Since first detection of
atmospheric neutrinos with energies $10^9-10^{11} eV$ in underground
neutrino experiments (Reines, et al., 1965; Achar, et al., 1965),
the target mass scale of the instruments - underground neutrino
telescopes - has grown to $\sim 5 \times 10^4$ tons
(Super-Kamiokande). In fact neutrino telescopes have become
universal instruments for investigation of the microworld (eg., in
the search for proton decay), as well as searching for cosmic
objects (eg., solar neutrinos, supernova neutrinos, WIMPS,
monopoles, etc.).

However to search for ultra-high energy (UHE, greater than $10^{15}
eV$) astrophysical neutrino sources, sites of the most energetic
processes (accelerators) in the Universe, neutrino telescopes of
effective detection volumes a cubic kilometer or more (KM3) are
necessary (Learned, 1993). Several prototype deep underwater optical
neutrino telescopes of $10^7$ $m^3$ in Lake Baikal, in the
Mediterranean Sea (NESTOR) and in the deep polar ice (AMANDA in the
Antarctic) are under construction now (see Gaisser et al., 1995 and
references therein).  A KM3 optical neutrino telescope could be
developed employing advances in technology for deep underwater
Cherenkov detectors (Chaloupka, et al., 1996) on a few year timescale.

Neutrino telescopes with a target scale greater than 1 km$^3$ have
also been proposed for studies of the upper boundary of the energy
spectrum of cosmic neutrinos. For example, if elementary particles
of maximal (Planck) masses $10^{24}- 10^{28} eV$ do exist, their
interactions and decays could produce neutrinos (and other
particles) of superhigh energies (SHE), up to $10^{24}-10^{28} eV$
(Markov and Zheleznykh, 1980).  Speculations upon the unknown source
of the observed highest energy cosmic rays have provided a plethora
of current models, many of which inevitably include significant or
even dominant neutrino fluxes.  For example, in the framework of
some GUT models with X-particle masses of $10^{23} eV$ and in
topological defect (TD) models, calculations of cosmic neutrino
fluxes with energies up to $10^{23} eV$ have been performed (Sigl ,
1996).  There is thus more motivation than in the past to search for
neutrinos of energies at and beyond the end of the observed cosmic
ray spectrum.

\begin{wrapfigure}[19]{r}{10.5cm}
\begin{picture}(260,180)(0,90)
\epsfig{file=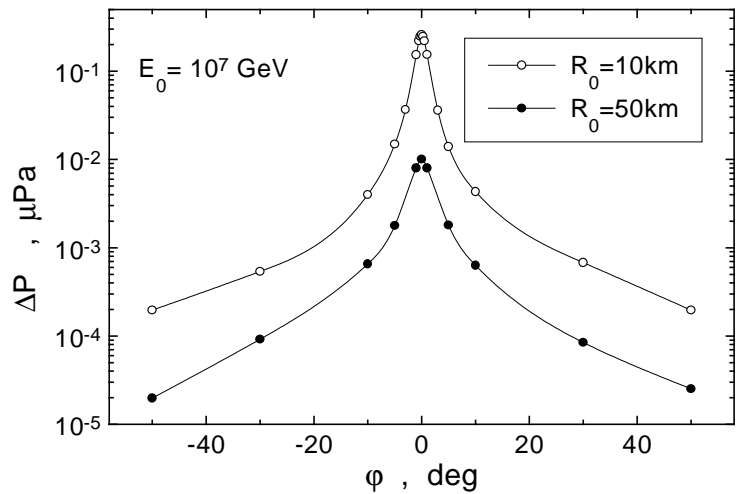,width=7.5cm,angle=0}
\end{picture}
\caption{\it The amplitude of an acoustic signal emitted by a
cascade with energy $10^7 GeV$ at distances of 10 and 50 km and at
large angles. For the perpendicular direction to the cascade axes
$\varphi$ = 0.}
\end{wrapfigure}

While until the present time the optical detection method remains
dominant, several alternative methods for studying UHE neutrino
interactions have been discussed and studied, both in the laboratory
and in the field during last two decades.  The two most studied
employ deep underwater acoustics (Askaryan and Dolgoshein, 1976;
Bowen, 1977; Learned, 1979) and radio wave neutrino detection in
cold Antarctic ice (Gusev and Zheleznykh, 1983; Markov and
Zheleznykh, 1986; Boldyrev et al., 1991; Ralston and McKay, 1990;
Zas et al., 1991, 1992).  The energy thresholds for radio and
acoustic detection are a few orders of value magnitude greater than
$10-50 GeV$ threshold typical for the deep underwater optical
detection. The threshold energy for optical detectors is one of
economics (above an MeV or so), whereas the limitation on radio and
acoustic detection is largely one of inherent signal-to-noise on
earth.  However the target mass scale of such alternative detectors
can be a few orders of magnitude greater than that of optical
detectors because the fundamental signal attenuation length in the
medium is given by tens of km for the acoustic signals, compared to
less than one hundred meters for light.

\section{INITIAL SADCO FOCUS: TO SEARCH FOR SHE NEUTRINOS}
Bipolar acoustic pulse production arises from the rapid
expansion of a region of material traversed by a neutrino
induced particle cascade, which ionizes and heats the
medium. Unfortunately it is an inefficient process, transferring only
a tiny fraction (order of $10^{-9}$) of the neutrino energy to
acoustic radiation (largely due to the smallness of the bulk
coefficient of thermal expansion).  An acoustic signal is emitted by
a neutrino induced cascade mainly in the direction perpendicular to
the cascade axis in a rather narrow solid angle. The angular
thickness of the disk is determined by the transverse size of the
effective emitting region (which for high energy particle cascades
in water is of the order of a few cm) divided the cascade length
(roughly ten meters), or a few milliradians. The initial spectrum
peaks at a few tens of kHz.

We have performed updated calculations of the acoustic pulse
production and propogation in the ocean, the details of which we
cannot present here because of space limitations.  The signal
spectrum at large angles and at large distances shifts to a lower
frequency region of a few kHz (Figures 1 and 2). Lower frequencies,
while costing dearly in amplitude, provide potentially dramatic
gains in solid angle and range over which events may be observed.
If SADCO could detect 1 kHz signals produced by SHE cascades at
distances of 10 to 50 km its effective detection volume could be
tens to hundreds of cubic kilometers.  The key to success is antenna
gain.

While acoustic detection of elementary particles was suggested by
G. A. Askaryan in the 1950's, well ahead of practical application,
it was not until twenty years later that the possibility of
construction of a deep ocean acoustic detector to study UHE
neutrinos was widely discussed.
\begin{wrapfigure}[18]{r}{10.5cm}
\begin{picture}(260,195)(10,100)
\epsfig{file=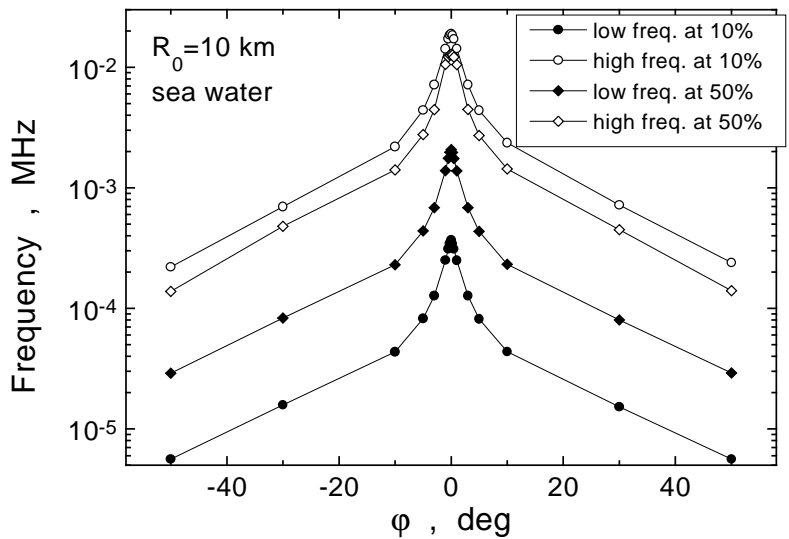,width=7.5cm,angle=0}
\end{picture}
\caption{\it The angular dependence of acoustic signal boundary
frequencies at distance of 10 km.}
\end{wrapfigure}
Even then it was not pursued because
at that time the threshold was thought too high and there were no
plausable sources of UHE neutrinos.  It was thought safer to start
with optical detectors which had guaranteed cosmic ray neutrino
signals in the $GeV$ energy range.  More recently with the advent of
larger optical neutrino detectors, and increased interest in UHE
neutrinos (as from AGNs), a deep underwater neutrino telescope SADCO
with threshold energy above $5 PeV$ was proposed to be deployed at a
depth of 3.5-4 km in the Ionean Sea near Pylos, Greece, at the site
of the NESTOR optical neutrino telescope (Karaevsky et al., 1993;
Dedenko et al., 1995). The sensitive volume of the SADCO neutrino
telescope would be greater than $10^8~m^3$. With the most optimistic
models, dozens of events per year might be seen from the Glashow
resonant neutrino interaction ($6.4~PeV$). But the main attraction
of SADCO is its ability to search for the SHE neutrino interactions
($E \geq 10^{17}eV$) in a huge water volume (Butkevich et al.,
1996).  The search for TD neutrinos should thus be one of goals for
SADCO if the registration volume reaches to hundreds cubic km.

\section{KAMCHATKA ARRAY AS A SADCO TEST BASE IN THE OCEAN}
It is quite attractive to consider the use of already existing
stationary sonar facilities, such as that placed in Kamchatka
region, as an acoustic detector of neutrinos.  This sonar has a
large plane phased-array, with 2400 hydrophones.  The array is
installed on the sea shelf and connected with on-shore equipment by
cable. The sector of view is 120$^{\circ}$.  The angular resolution
in the horizontal plane is 0.8$^{\circ}$ in each of 150 (virtual)
parallel fan-shaped beams. The vertical angular width is
7$^{\circ}$. The gain of this array is 2500 at 1400 Hz.

One can approximately estimate the detection capability of such a
system.  Let us assume that a neutrino induces a cascade with an
energy $E = 10^{20} eV$. An acoustic impulse with time duration of
25$\mu$sec is generated. The maximum of the acoustic pressure of
such a signal at a range of 10 kilometers would be 2000$\mu$Pa. The
level of acoustic pressure at a frequency of 1400 Hz is however only
approximately 19$\mu$Pa.  Because the mean amplitude of sea noise in
this frequency range is 100-200$\mu$Pa (Beaufort Number 0-6), the
signal/noise ratio at the array would be 0.2-0.01. However, after
beamforming (antenna gain) the output signal/noise ratio should be
10-0.5.  In other words such stationary sonar can detect cosmic
particles with energy more than $10^{20} eV$ and in a very large
volume, which we conservatively estimate to be more than 100 cubic
kilometers.

Acoustic signals emitted over a large solid angle in the ocean
generally are complex when observed from some distance due to the
guiding of sound around the depth of sound velocity minimum (the
SOFAR zone).  The small solid angle of peak emission from neutrino
events may thus be employed to advantage in discriminating against
background.  Despite the fact that the frequency range of the
existing sonar is not optimal (we would like higher bandwidth) the
very large detection volume can compensate for the frequency deficit
and greatly increases the detection probability of extremely
energetic particles.

From the practical standpoint, the economic benefit of such an
exploration is evident, because this sonar is in operation. Our
suggested variant of acoustic detection of elementary particles can
thus be very suitable for preliminary search experiments. Moreover
this permits the development of different methodologies and
algorithms for the detection and recognition of neutrino events. It
is expedient to construct a special acoustical transducer suspended
from a ship for simulation of neutrino impulse signals with
different space-time parameters and calibrate the system. Little
else is needed in terms of hardware.

\section{CONCLUSIONS}
We have outlined a program for study of acoustic detection of
neutrinos in the oceans employing existing sonar arrays. This
program, which requires almost no hardware investment, can begin as
soon as support is available for personnel. First it will permit
development of detection techniques in the real ocean, something not
heretofore available to neutrino physicists.  Moreover, the existing
equipment should permit a useful physics exploration for neutrinos
of extreme energies, not far above energies of cosmic rays already
observed by extensive air shower arrays. We look forward to tests of
this system with a year.

\section{ACKNOWLEDGEMENTS}
One of the authors (I.Zh.) thanks the Russian Foundation of Basic
Research for support (grant 96-02-18594).

\section{REFERENCES}
\setlength{\parindent}{-5mm}
\begin{list}{}{\topsep 0pt \partopsep 0pt \itemsep 0pt \leftmargin 5mm
\parsep 0pt \itemindent -5mm}
\vspace{-15pt}
\item Achar, C. V. et al., {\it Phys. Lett.}, 18, 196 (1965).
\item Askaryan, G. A. and Dolgoshein, B. A., 1976 DUMAND Summer
Workshop, Hawaii.
\item Boldyrev, I. N. et al., {\it Proc. 3rd Int. Workshop on
Neutrino Telescope}, ed. Milla Baldo Ceolin, Venezia, 337 (1991).
\item Bowen, T., {\it Conference Papers, 15th ICRC}, Plovdiv, 6, 277
(1977).
\item Butkevich, A. V. et al., {\it Proc. Int. School Particles and
Cosmology (1995)}, Baksan, World Scientific, 306 (1996).
\item Chaloupka, V. et al.,{\it LBL - 38321}, Feb. 1996.
\item Dedenko, L. G. et al., {\it Proc. 24th ICRC}, Rome, 1, 797 (1995).
\item Gaisser, T. K., Halzen, F , and Stanev, T., {\it Phys. Rep.},
258, 173 (1995).
\item Gusev, G. A. and Zheleznykh, I. M., {\it Pisma
Zh. Exp. Teor. Fiz.}, 38, 505 (1983).
\item Karaevsky, S. Kh. et al., {\it Proc. 23rd ICRC}, Calgary, 4,
550 (1993).
\item Learned, J. G., {\it Phys. Rev. D},19, 3293 (1979).
\item Learned, J. G., {\it Nucl. Phys. B (Proc. Suppl.)}, 33 A,B, 77
(1993).
\item Markov, M. A. and Zheleznykh, I. M., in {\it Proc 1979 Dumand
Summer Workshop at Khabarovsk and Lake Baikal}, ed. J.Learned, p.177
(1980).
\item Markov, M. A., Zheleznykh, I.M., {\it Nucl. Inst. Methods}, A
248, 242 (1986).
\item Ralston, J. P. and McKay, D. W., {\it Nucl. Phys. B
(Proc. Suppl.)}, 14A, 356 (1990).
\item Reines, F. et al, {\it Phys. Rev. Lett.}, 15, 429 (1965);
\item Sigl, G., talk at {\it HENA Workshop}, 26 May - 9 June 1996,
Aspen.
\item Zas, E., Halzen, F., and Stanev, T., {\it Phys. Lett. B}, 257,
432 (1991); {\it Phys. Rev. D}, 45, 362 (1992).
\end{list}
\end{document}